\documentclass[aps,prl,reprint,twocolumn,showpacs,superscriptaddress,longbibliography,10pt]{revtex4-1}
\usepackage[english]{babel}
\usepackage{graphicx}
\usepackage{bm}
\usepackage{color}
\usepackage{amsmath}
\usepackage{amsfonts}
\usepackage{epstopdf}

\newcommand{\soglia}{\textrm{th}}

\begin{document}

\author{G. Bertaina}
\affiliation{Dipartimento di Fisica, Universit\`a degli Studi di Milano, via Celoria 16, I-20133 Milano, Italy}

\author{M. Motta}
\affiliation{Department of Physics, The College of William and Mary, Williamsburg, Virginia 23187, USA}

\author{M. Rossi}
\affiliation{Scuola Normale Superiore, Piazza dei Cavalieri 7, I-56126 Pisa, Italy}
\affiliation{International Center for Theoretical Physics (ICTP), Strada Costiera 11, I-34154 Trieste, Italy}
\affiliation{Dipartimento di Fisica e Astronomia, Universit\`a degli Studi di Padova, via Marzolo 8, I-35131 Padova, Italy}

\author{E. Vitali}
\affiliation{Department of Physics, The College of William and Mary, Williamsburg, Virginia 23187, USA}

\author{D.E. Galli}
\affiliation{Dipartimento di Fisica, Universit\`a degli Studi di Milano, via Celoria 16, I-20133 Milano, Italy}

\title{One-dimensional liquid $^4$He: dynamical properties beyond Luttinger liquid theory} 

\begin{abstract}
We compute the zero-temperature dynamical structure factor of one-dimensional liquid $^4$He by means of state-of-the-art 
Quantum Monte Carlo and analytic continuation techniques.
By increasing the density, the dynamical structure factor reveals a transition from a highly compressible critical liquid to a quasi-solid regime.
In the low-energy limit, the dynamical structure factor can be described by the quantum hydrodynamic Luttinger liquid theory, with a Luttinger parameter spanning all possible values by increasing the density.
At higher energies, our approach provides quantitative results beyond the Luttinger liquid theory. In particular, as the
density increases, the interplay between dimensionality and interaction makes the dynamical structure factor manifest
a pseudo {\it{particle-hole}} continuum typical of fermionic systems. At the low-energy boundary of such region and moderate densities, we find consistency, within statistical uncertainties, with predictions of a power-law structure by the recently-developed 
non-linear Luttinger liquid theory.  
In the quasi-solid regime we observe a novel behavior at intermediate momenta, which can be described by new analytical relations that we derive for the hard-rods model.
\end{abstract}

\maketitle 

One-dimensional (1D) quantum systems exhibit some of the most diverse and fascinating phenomena of condensed matter Physics \cite{giamarchi,cazalilla_one_2011,imambekov_one_2012}.
Among the most spectacular signatures of the interplay between quantum fluctuations, interaction and reduced
dimensionality, are the breakdown of ordered phases in presence of short-range interactions \cite{mermin_absence_1966}, and the 
loosened distinction between Bose and Fermi behavior \cite{girardeau_relationship_1960}.
The study of quasi-1D quantum systems is a very active research field, aroused by the experimental investigation 
of electronic transport properties \cite{chang_1996,yao_1999,bockrath_1999,aleshin_2004,zotov_2000}, by the fabrication of long 1D arrays of Josephson junctions \cite{chow_length_1998}, and recently 
corroborated by the availability of ultracold atomic gases in highly anisotropic traps and optical lattices 
\cite{bloch_many_2008,cazalilla_one_2011,fabbri_dynamical_2015,meinert_probing_2015}, as well as by experiments on confined He atoms 
\cite{yager_nmr_2013,savard_hydrodynamics_2011,taniguchi_dynamical_2013,vekhov_mass_2012,vekhov_2014}.

The low-energy properties of a wide class of Bose and Fermi 1D quantum systems \cite{vignale,giamarchi} are notoriously
captured by the phenomenological
Tomonaga-Luttinger liquid (TLL) theory \cite{Tomonaga_1950,luttinger_exactly_1963,haldane_effective_1981},
characterized by collective phonon-like excitations.
This theory introduces two conjugate Bose fields $\phi(x)$, $\theta(x)$ describing, respectively, the density and phase fluctuations of the 
field operator $\psi(x) = \sqrt{ \rho + \partial_x\phi(x) } \, e^{i \theta(x)}$, where $\rho$ is the average density. 
Those fields are described by the exactly-solvable low-energy effective Hamiltonian:
\begin{equation}
H_{LL} = \frac{\hbar}{2\pi} \int dx \, \left( c K_L \partial_x\theta(x)^2 + \frac{c}{K_L} \, \partial_x\phi(x)^2 \right) \quad .
\end{equation}
Although in general the TLL parameter $K_L$ and the sound velocity $c$ are independent quantities (notably in lattice models), for Galilean-invariant systems $c=\frac{v_F}{K_L}$ \cite{haldane_effective_1981}, $v_F=\frac{\hbar k_F}{m}$ being the Fermi velocity and $k_F = \pi \rho$ the Fermi wavevector of a 1D ideal Fermi gas (IFG), and $K_L$ is thus related to the compressibility 
$\kappa_S$ by $m \, K_L^2 = \hbar^2 \pi^2 \rho^3 \kappa_S$.
Such collective excitations are revealed by the low-momentum and low-energy behavior of the dynamical structure factor:
\begin{equation}
\label{eq:sofk}
S(q,\omega) = \int dt \frac{e^{i \omega t}}{2\pi N} \langle e^{\frac{itH}{\hbar}}\rho_q e^{-\frac{itH}{\hbar}} \rho_{-q} \rangle \quad , 
\end{equation}
where $\rho_q = \sum_{i=1}^N e^{i q x_i}$ is the Fourier transform of the density operator, $N$ the number of particles, 
$H$ the Hamiltonian and $x_i$ the position of the i-th particle \cite{[{Such quantity is related to the imaginary part of the density-density response function by the fluctuation-dissipation theorem. See }][{}]kubo_fluctuation_1966}. 
A complete characterization of density fluctuations requires to compute \eqref{eq:sofk} also beyond the limits of applicability of TLL theory.
A deep insight in the characterization of \eqref{eq:sofk} at higher frequencies is provided by the phenomenological nonlinear TLL theory
\cite{imambekov_phenomenology_2009,imambekov_one_2012}; for integrable models, quantitative results are also provided by nonperturbative numeric 
calculations \cite{caux_dynamical_2006,mourigal_fractional_2013,lake_multispinon_2013,fabbri_dynamical_2015,meinert_probing_2015}.
For physically-relevant non-integrable systems, on the other hand, the study of \eqref{eq:sofk} requires more general approaches.

In this Letter, we probe the excitations of 1D liquid $^4$He by evaluating its complete zero-temperature dynamical structure factor with fully {\em{ab-initio}} methods.
When strictly confined in 1D, $^4$He provides a spectacular condensed-matter realization of a TLL, having the unique 
feature of spanning all possible values of $K_L$ by only varying the density.
The interest in this system emerges also in connection with experimental realizations and theoretical characterizations of
quasi-1D He systems confined inside nanopores \cite{kresge_ordered_1992,delmaestro_2010,delmaestro_he4_2011,taniguchi_dynamical_2013} or moving inside dislocation lines in 
crystalline He samples \cite{boninsegni_screw_2007,vekhov_mass_2012,vekhov_2014}.
A realistic microscopic description of the system is provided by the Hamiltonian:
\begin{equation}
\label{eq:ham}
H = - \frac{\hbar^2}{2m} \sum_{i=1}^N \frac{\partial^2}{\partial x_i^2} + \sum_{i<j=1}^N V(x_i-x_j) \quad ,
\end{equation}
$V(x)$ being the well-established Aziz potential \cite{aziz_1979}.
We access $S(q,\omega)$ by performing an inverse Laplace transform of the imaginary-time correlation function:
\begin{equation}
\label{eq:laplace}
F(q,\tau) = \frac{1}{N} \langle e^{\frac{\tau H}{\hbar}} \rho_q e^{-\frac{\tau H}{\hbar}} \rho_{-q} \rangle = \int_0^\infty d\omega e^{-\tau\omega} S(q,\omega).
\end{equation}
We compute $F(q,\tau)$ using the Path Integral Ground State (PIGS) method \cite{sarsa_2000,spigs_2003}, which provides unbiased \cite{potatoes} 
estimates of ground-state properties and imaginary-time correlations by statistically sampling the wavefunction $\Psi_\tau = 
e^{-\tau H} \Psi_T$, where $\Psi_T$ is a trial state \cite{reatto_1967,swf_1988}, non-orthogonal to the ground state of $H$.  
At sufficiently large $\tau$, the expectation values over $\Psi_\tau$ are compatible with ground-state averages. 
We simulate up to $N=160$ particles using periodic boundary conditions and find that our results are representative of the thermodynamic limit already for $N=40$ particles within statistical uncertainty (see Supplemental Material \cite{supplemental}).
Inverting the Laplace transform in Eq. \eqref{eq:laplace} is notoriously an ill-posed inverse problem, meaning that many possible $S(q,\omega)$ are compatible with the imaginary-time data.
However, a number of inversion strategies have provided reliable results for physically relevant systems \cite{sandvik_stochastic_1998,mishchenko_diagrammatic_2000,reichman_analytic_2009,gift_2010}. In this Letter, we use the state-of-the-art Genetic Inversion via Falsification of Theories (GIFT) algorithm \cite{gift_2010,overpressurized,saccani_bose_2011,saccani_excitation_2012,nava_superfluid_2012,he3_dsf,arrigoni_excitation_2013,rota_2013}. 

\begin{figure}[ptb]
\begin{center}
\includegraphics[width=8.5cm]{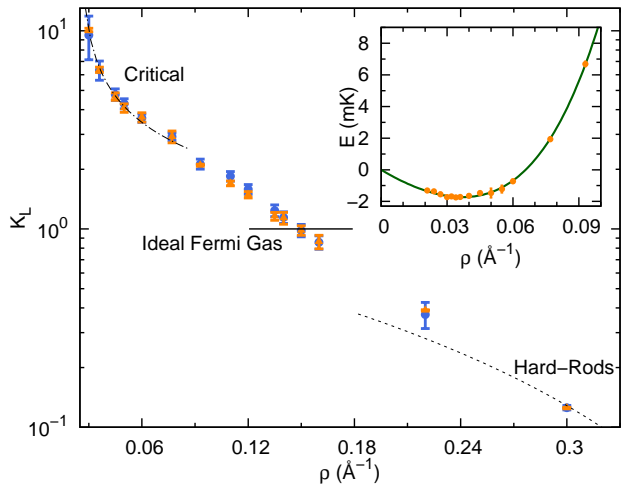}
\caption{(color online) 
TLL parameter $K_L$, from the compressibility $\kappa_S^{-1} = 
\rho \partial_\rho \left( \rho^2 \partial_\rho E(\rho) \right)$ (blue circles) and the low-$q$ behavior 
of $S(q)$ (orange triangles). Superimposed lines are described in the text. Inset: equation of state $E(\rho)$.
}
\label{fig:klut}
\end{center}
\end{figure}

\begin{figure}[thp]
\begin{center}
\includegraphics[width=8.5cm]{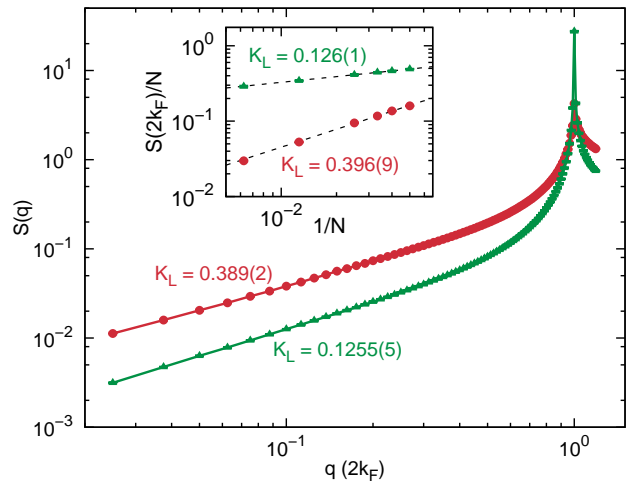}
\caption{(color online) Static structure factor $S(q)$ at $\rho = 0.22$, $0.30$ \AA$^{-1}$ (red circles, green triangles).
         Inset: Scaling of $S(2 k_F)$ with $N$ at the same densities (dashed lines: fit to a power-law). 
         Values of $K_L$ from $c$ and the scaling of $S(2 k_F)$ are reported.
}
\label{fig:skscaling}
\end{center}
\end{figure}

We study the Galilean-invariant liquid phase which is notoriously stable above the density $\rho_{sp} = 0.026(2)$~\AA$^{-1}$, where it 
undergoes a spinodal decomposition \cite{stan_interstitial_1998,krotscheck_properties_1999,boninsegni_ground_2000}, namely the formation of liquid droplets. In Fig.~\ref{fig:klut}, we compute the TLL parameter $K_L$ of the system as a function of $\rho > \rho_{sp}$ 
from both the compressibility and the sound velocity, inferred from the low-momentum behavior of the static
structure factor $S(q)=F(q,0) \simeq K_L \frac{q}{2k_F}$.
The good agreement between the two estimates over the whole density range confirms their accuracy, and the internal consistency of our approach. 
Close to the spinodal decomposition, the sound velocity provides a more precise estimate of $K_L$ \footnote{Estimating $K_L$ from $\kappa_S$ requires differentiation of the equation of state, which is fitted to a polynomial. Uncertainties on the fit parameters propagate to $\kappa_S$, resulting in unavoidably large error bars near the spinodal decomposition, where the compressibility diverges.}.
As the density increases, $K_L$ monotonically decreases
from $\infty$ to $0$, manifesting three fundamental regimes.
At density $\rho \lesssim 0.06$~\AA$^{-1}$ the system is in the spinodal critical regime and we observe $K_L \propto (\rho-\rho_{sp})^{-\zeta}$ with $\zeta \simeq 0.5$.
This is equivalent to a dependence $c\propto (P-P_{sp})^{\nu}$ of sound velocity with the pressure difference $P-P_{sp}$, with $P_{sp}$ the pressure at the spinodal point and $\nu=\zeta/(2\zeta+1)\simeq 0.25$, which is interestingly consistent with the critical value in three-dimensional helium \cite{albergamo_phonon-roton_2004,solis_liquid_1992,boronat_monte_1994,bauer_path-integral_2000}.
At density $\rho \gtrsim 0.30$~\AA$^{-1}$ we observe instead a good agreement with the hard-rods (HR) model \cite{mazzanti_ground_2008}, defined by $V(x) = \infty$ for $|x|<a$ and $0$ otherwise. In Fig.~\ref{fig:klut} we take $a = 2.139\text{\AA}$, which is the scattering length of the repulsive part of the $^4$He potential as in \cite{[{See }][{. Note that due to the hard core, the 1D scattering problem has the same boundary condition as the 3D reduced radial solution.}]kalos_1974}. 
The HR model spans all values of $K_L=(1 - \rho a)^2 < 1$ as a function of the density.
At the intermediate density $\rho \simeq 0.150$~\AA$^{-1}$ $^4$He attains $K_L=1$, which is the TLL parameter of the Tonks-Girardeau gas of impenetrable point-like Bosons \cite{girardeau_relationship_1960} and of the 1D IFG.

\begin{figure*}[th]
\begin{center}
{
\includegraphics[width=\textwidth]{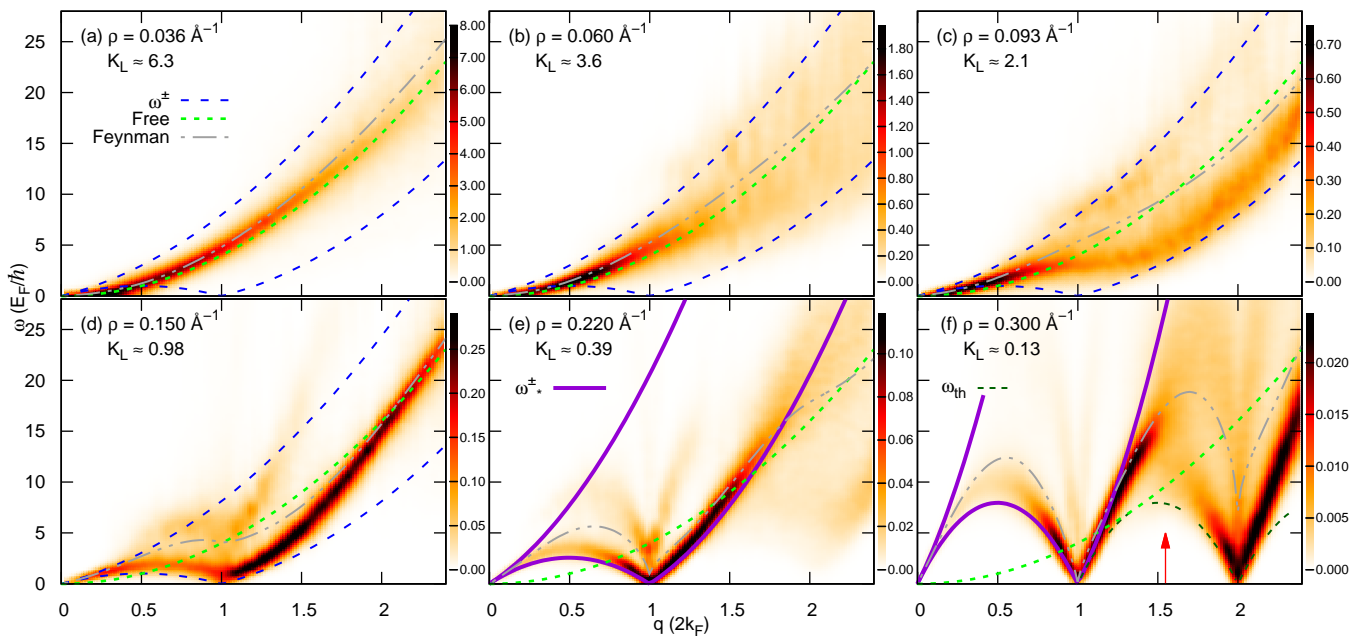}
\caption{(color online) Color plot of $S(q,\omega)$ at several densities and corresponding $K_L$. 
Feynman approximation $\omega_F(q)$ (gray dash-dotted lines) and the free particle dispersion $\hbar q^2/2m$ (green dotted lines) are drawn for comparison.
Panels (a-d) show also the bounds $\omega^{\pm}(q)$ of 
the particle-hole band (blue dashed line), while panels (e-f) show the bounds $\omega^{\pm}_*(q)$ of the HR elementary excitations (violet solid line).
Panel (f) shows the low-energy threshold $\omega_{\soglia}(q)$ of HR with $K_L = 0.125$ (double-dashed line), and momentum ${\cal Q}_1$ (red arrow).
Values of $S(q,\omega)$ beyond scale are plot in black.
}
\label{fig:spectra}
}
\end{center}
\end{figure*}

The diverse behavior of $^4$He is a peculiar consequence of the interplay between the hard-core repulsion and the Van der Waals 
attraction in the interaction potential, and the mass of the atoms.
It has been recently recognized that the TLL parameter of $^3$He features a similar high-density behavior \cite{astrakharchik_luttinger_2014}; 
the low-density behavior, however, is remarkably different as the smaller mass of $^3$He prevents a spinodal decomposition, maintaining $K_L$ and the compressibility below a finite value.

In view of the universality of TLL theory, knowledge of $K_L$ sheds light on the low-momentum and low-energy 
structure of $S(q,\omega)$.
TLL theory also predicts \cite{luther_single_1974,castro_neto_1994,astrakharchik_motion_2004} a power-law singularity $S(q = 2k_F j, \omega) \sim \omega^{2(j^2K_L - 1)}$ for $\omega \to 0$  and integer ($j\in\mathbb{N}$) multiples of $2 k_F$. Such singularity is strictly related to the emergence of quasi-Bragg peaks in the static structure 
factor, featuring a sub-linear growth $S(2 k_Fj) \propto N^{1 - 2 j^2 K_L}$ \cite{mazzanti_ground_2008} with the number of particles.
The height of the $j$-th peak diverges, in the thermodynamic limit, provided that $2j^2K_L < 1$. 
In Fig. \ref{fig:skscaling} we observe the emergence of quasi-Bragg peaks in $S(2 k_F)$ at densities $\rho > 0.196(5)$ 
\AA$^{-1}$, where $K_L < 1/2$. This is naturally expected since the small compressibility sets up a diagonal quasi-long range 
order, while crystallization is prohibited by the dimensionality and by the range of the interaction \cite{mazzanti_ground_2008}.
The scaling of $S(2 k_F)$ with $N$, reported in the inset of Fig.~\ref{fig:skscaling}, provides an alternative
estimate of $K_L$, in agreement with the results in Fig.~\ref{fig:klut}.

The rich physical behavior suggested by the TLL parameter is notably unveiled by the dynamical 
structure factor, that our approach characterizes over the 
entire momentum-energy plane.
Fig.~\ref{fig:spectra} shows $S(q,\omega)$ as a function of momentum and frequency, in Fermi units $2k_F$ and $E_F/\hbar=\hbar k_F^2/2 m$ respectively, at several 
representative densities. We show also Feynman's approximation for the excitation spectrum $\omega_F(q) = \hbar q^2/2mS(q)$, which postulates a single mode saturating the f-sum rule $\hbar q^2/2m = \int d\omega S(q,\omega)\omega$. Departures from the Feynman spectrum indicate a broadening or the presence of multiple modes \cite{[{A pioneering, but less general, fit of imaginary-time density correlations was performed for the dipolar gas in }][{}]depalo_low_2008}.

As expected, for small $q$ and $\omega$, $S(q,\omega)$ is always peaked around the phonon dispersion
relation $\omega = c q$.
On the other hand, the high-energy scenario is strikingly different and strongly dependent on the 
density.
At $K_L \simeq 6.3$ (Fig.~\ref{fig:spectra}a) the spectral weight is very close to the free particle dispersion, consistently with similar predictions for 3D helium at negative pressures \cite{albergamo_phonon-roton_2004,solis_liquid_1992,boronat_monte_1994,bauer_path-integral_2000}.
Such behavior is common to the Lieb-Liniger contact interaction model at large $K_L$ \cite{lieb_exact_1963a,lieb_exact_1963,caux_dynamical_2006}, although in the case of $^4$He the physical origin of such a behavior lies in the spinodal critical point.
At large momentum ($q \gtrsim k_F$) and energy we observe a broadening of $S(q,\omega)$, that makes more and more 
pronounced as $K_L$ decreases (Fig.~\ref{fig:spectra}b,c). 
As in the Lieb-Liniger model \cite{caux_dynamical_2006}, the spectral weight of $S(q,\omega)$ partially fills the 
particle-hole band of the 1D IFG, enclosed between the dispersion relations $\omega^\pm(q) = 
\left| v_F q \pm \hbar q^2/2m\right|$.
In both cases, this reveals a tendency for fermionization \cite{girardeau_relationship_1960}: the repulsive 
interaction between 1D bosons mimics the Pauli exclusion principle, and makes $S(q,\omega)$
manifest the particle-hole continuum typical of spinless free fermions.
At $K_L \simeq 2.1$ (Fig.~\ref{fig:spectra}c) the spectral weight of $^4$He starts to concentrate again, emerging as a phonon and then 
bending downwards to approach $\omega^-(q)$. Such peculiar behavior is reminiscent of the 
deflection of the Bogoliubov mode in 3D systems of hard spheres \cite{rota_2013,rossi_path_2013}, with the notable difference that in 1D the spectral weight at $q\simeq 2k_Fj$ is non-zero up to very low frequency.
At $K_L \simeq 1$ (Fig.~\ref{fig:spectra}d) the incipient concentration of the spectral weight makes strikingly manifest and takes place around a low-energy excitation, which is close to $\omega^-(q)$ 
for $q < 2 k_F$ and approaches the free particle dispersion relation for higher momentum. 
However, $S(2 k_F,\omega)$ is almost flat at low frequency $\omega \lesssim E_F/\hbar$, within our resolution (see Supplemental Material \cite{supplemental}), analogously to the Tonks-Girardeau and IFG models.
Above the low-energy excitation a lower-intensity secondary structure overhangs; for 
$K_L < 1$ (Fig.~\ref{fig:spectra}e,f) it evolves into a well-defined high-energy 
structure attaining a non-zero local minimum at $q= 2k_F$, in correspondence of the 
free-particle energy. Although a precise characterization of this structure requires further investigation, it is reminiscent of a 3D rotonic behavior or of multi-phonons \cite{cowley_inelastic_1971,galli_rotons_1996,rota_2013,rossi_path_2013}.
For $K_L \simeq 0.39$ (Fig.~\ref{fig:spectra}e) $S(q,\omega)$ is mostly distributed in a region with boundaries $\omega^\pm_*(q)$, which are 
modified with respect to $\omega^\pm(q)$ as an effect of interaction, and the spectral weight concentrates 
close to the lower branch $\omega^-_*(q)$. We notice that $\omega^\pm_*(q) = \omega^\pm(q) /K_L$ (solid lines in Fig.~\ref{fig:spectra}e,f). A similar behavior can be discerned \cite{[{We analyzed data from }][{}]panfil_metastable_2013} in the Super Tonks-Girardeau gas \cite{astrakharchik_tonks_2005,
haller_realization_2009}, a gaseous excited state of the attractive Lieb-Liniger model. 
This behavior can be quantitatively explained: in the high-density regime the main interaction effect is volume exclusion, as in the HR model. The solution of such model via the Bethe Ansatz technique \cite{nagamiya_statistical_1940,sutherland_quantum_1971,noi_HR_2016} shows 
that the eigenfunctions of the HR Hamiltonian can be mapped onto those of an IFG with increased
density $\rho/(1-\rho a)$, thus yielding a scaling factor $(1-\rho a)^{-2}=K_L^{-1}$ in the boundaries of the particle-hole band.

The distribution of spectral weight changes dramatically for $K_L \simeq 0.125$ (Fig.~\ref{fig:spectra}f) for $2k_F < q < 4k_F$, where the low-energy excitation rapidly broadens and flattens at $q \simeq
3.2 k_F$, and concentrates again at a lower energy around $q \simeq 4k_F$.
A quantitative explanation of this effect can be given in the light of the recently-developed nonlinear
TLL theory \cite{imambekov_one_2012}, again modeling ${}^4$He atoms with HR. Nonlinear TLL theory assumes 
the existence of a low-energy threshold $\omega_{\soglia}(q)$, below which no excitations are present.
Interpreting an excitation with frequency $\omega \gtrsim \omega_{\soglia}(q)$ as the creation of a mobile 
impurity in an otherwise usual TLL, nonlinear TLL theory shows that $S(q,\omega)$ features a power-law
singularity:
\begin{equation}
\label{eq:nltll}
S(q,\omega) \propto \Theta\left( \omega - \omega_{\soglia}(q) \right) \, |\omega-\omega_{\soglia}(q)|^{-\lambda(q)}\,,
\end{equation}
where $\lambda(q)$ is a function of $K_L$ and $\omega_{\soglia}(q)$ \cite{imambekov_phenomenology_2009} and $\Theta(\omega)$ is the Heaviside step function. 
The expansion $\omega_{\soglia}(q) \approx c q - \hbar q^2/2m^*$ of the low-energy threshold around $q=0$ 
defines the effective mass $m^*$, which sets the energy scale where modifications from TLL theory take 
place \cite{imambekov_phenomenology_2009}. The effective mass is a function $1/m^* = c \, \partial_\mu\left( \, 
c \, \sqrt{K_L} \, \right) / K_L$ of $K_L$ and the chemical potential $\mu$ \cite{pereira_dynamical_2006,
imambekov_phenomenology_2009}. For the HR model we indeed derive $m/m^* = 1/K_L$, 
indicating that $\omega_{\soglia}(q) \approx \omega^-_*(q)$ for small momentum. This is again confirmed over the whole range 
$0\leq q \leq 2k_F$ by the analytical solution of the HR model \cite{noi_HR_2016}.
Away from this basic region, the low-energy threshold repeats periodically \cite{castro_neto_1994,
imambekov_one_2012,cherny_2012} as shown in Fig.~\ref{fig:spectra}f: therefore $\omega_{\soglia}(q)=
\omega^-_*(q-2n k_F)$ with $2 n k_F < q < 2(n+1) k_F$ and $n$ integer.

\begin{figure}[tbp]
\begin{center}
\includegraphics[width=8.5cm]{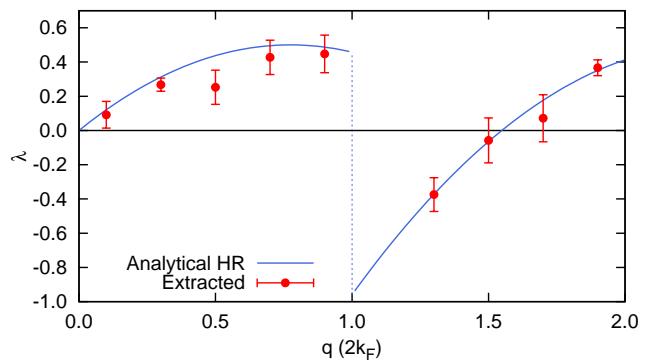}
\caption{(color online) Analytical non-linear TLL exponent Eq. \eqref{eq:lambda} for HR with $K_L=0.125$ (solid line) and PIGS+GIFT (circles) fitted exponents of $^4$He at density $\rho=0.3\text{\AA}^{-1}$.}
\label{fig:exp}
\end{center}
\end{figure}

For the HR model, given the analytic expressions of $K_L$ and $\omega_{\soglia}(q)$, we extract the exponents following \cite{imambekov_phenomenology_2009}: 
\begin{equation}
\lambda(q)=-2 \left(\tilde{q}-n\right)\left(\tilde{q}-n-1\right)\;,\qquad \tilde{q}\equiv q a/2\pi \;. \label{eq:lambda}
\end{equation}
In Fig.~\ref{fig:exp} we show $\lambda(q)$ for a HR system with the same $K_L$ as in Fig.~\ref{fig:spectra}f, comparing it to numerically extracted exponents as described below. $\lambda(q)$ is a piecewise continuous function of $q$, with jump singularities at $q=2nk_F$.
For $0\leq q < 2 k_F$, $\lambda(q)>0$ and $S(q,\omega)$ diverges close to $\omega_{\soglia}(q)$. After $q=2k_F$, $\lambda(q)$ changes sign and thus $S(q,\omega)$ vanishes close to $\omega_{\soglia}(q)$. In fact, for $2k_F < q \lesssim 3.2 k_F$, the spectral weight concentrates much above $\omega_{\soglia}(q)$, around $\omega^-_*(q)$, a feature which is even beyond nonlinear TLL theory.
Eq. \eqref{eq:lambda} predicts a flat $S(q,\omega)$ at the special wavevectors ${\cal Q}_n=2\pi n/a$, 
consistently with a previous result \cite{mazzanti_ground_2008} based on exact properties of 
the HR model.
We indeed observe almost flat $S(q,\omega)$ at ${\cal Q}_1 = 3.24 \, k_F \simeq 2\pi/a$ (red arrow in Fig. ~\ref{fig:spectra}f).
Beyond ${\cal Q}_1$ the divergence reappears, since $\lambda(q)<0$. 

To quantitatively verify prediction \eqref{eq:lambda}, for some momenta we have performed much more refined
reconstructions at $\rho=0.3\text{\AA}^{-1}$, imposing $S(q,\omega)=0$ \cite{sandvik_constrained_2015} below the exact $\omega_{\soglia}(q)$ for the HR model, 
and fitting the obtained spectrum to a power law (see Supplemental Material \cite{supplemental}).
The obtained exponents are indicated in Fig.~\ref{fig:exp}:
this procedure does not disprove the power-law model \eqref{eq:nltll} in a range of frequencies up to $\sim\omega_\soglia(q)+E_F/\hbar$, depending on momentum \footnote{Calculation of $\lambda(q)$ at momenta slightly larger than $2k_F$ and $4k_F$ was prevented by the difficulty of 
resolving a vanishing spectrum in a narrow frequency range below the dominant higher-energy peak.}, and yields exponents $\lambda(q)$ which are consistent with the nonlinear TLL prediction \eqref{eq:lambda} within statistical uncertainty. This result is quite remarkable, since no prior knowledge about $S(q,\omega)$ has been enforced in the analytic continuations, except for the f-sum rule and the exact threshold for HR \footnote{A small discrepancy is seen around $q=k_F$ where in fact the threshold for $^4$He seems to be higher than that predicted by the HR model.}.

We have thus provided a robust description of the system in the experimentally-relevant high-density regime, based on the HR model, which almost fully characterizes the spectrum at low and intermediate energies.
The novel structure predicted around momenta that are multiples of $2\pi/a$ is relevant, and would be very interesting to experimentally observe, for all quantum excluded-volume systems, such as liquid He inside nanopores, Rydberg gases \cite{schempp_full_2014,schauss_observation_2012} and Super-Tonks-Girardeau gases.

\begin{acknowledgments}
We acknowledge very useful discussions with G. Astrakharchik. We are grateful to A. Parola for revising the manuscript. We thank M. Panfil and co-authors for providing us with their data on the Super-Tonks-Girardeau gas.
The simulations were performed on the supercomputing facilities at 
CINECA and at the Physics Departments of the Universities of Milan and Padua. We thank the Computing Support Staff at INFN and Physics Department of the University of Milan.
We acknowledge the CINECA and the Regione Lombardia
award $\text{LI03p-UltraQMC}$, under the LISA initiative, for the availability of
high-performance computing resources and support.
M.M. acknowledges funding from the Dr. Davide Colosimo Award, celebrating
the memory of physicist Davide Colosimo. M.M. and E.V. acknowledge support from the Physics Department of the University of Milan, the Simons Foundation and NSF (Grant no. DMR-1409510). G.B. and D.E.G. acknowledge funding from D.E. Pini.
\end{acknowledgments}
 

%

\pagebreak
\onecolumngrid
\vspace{\columnsep}
\newpage
\begin{center}
\textbf{\large Supplemental Material: One-dimensional liquid $^4$He: dynamical properties beyond Luttinger liquid theory}
\end{center}
\vspace{2cm}
\twocolumngrid

\setcounter{equation}{0}
\setcounter{figure}{0}
\setcounter{table}{0}
\setcounter{page}{1}
\makeatletter
\renewcommand{\theequation}{S\arabic{equation}}
\renewcommand{\thefigure}{S\arabic{figure}}
\renewcommand{\bibnumfmt}[1]{[S#1]}
\renewcommand{\citenumfont}[1]{S#1}
\addtolength{\textfloatsep}{5mm}

Note: citations in this Supplemental Material refer to the bibliography in the main paper.

\section{Path-Integral Ground State method}

The Path Integral Ground State (PIGS) Monte Carlo method is a projector 
technique that provides direct access to ground-state expectation values
of bosonic systems, given the microscopic
Hamiltonian $\hat H$ [34]
. The method is exact, within unavoidable statistical error bars, which can nevertheless be reduced by performing longer simulations, as in all Monte Carlo methods.  Observables $\hat{O}$ are calculated as $\langle \hat{O}\rangle = \lim_{\tau\to\infty}\langle\Psi_\tau|\hat{O}|\Psi_\tau\rangle/\langle\Psi_\tau|\Psi_\tau\rangle$, 
where $\Psi_\tau = e^{-\tau \hat{H}} \Psi_T$ is the imaginary-time projection of an initial trial wave-function $\Psi_T$. Provided non-orthogonality to the ground state, the quality of the wave-function only influences the projection time practically involved in the limit and the variance of the results.

In our study we have employed a trial Shadow wave function (SWF) [38]
, which is known to provide a very accurate description of 
the ground state of liquid and solid $^4$He in higher dimensions, since
it introduces high order correlations in an implicit way by means of 
auxiliary variables. The SWF has the form $\Psi_T({\bf{R}}) = \int {\bf{dS}}\;  G({\bf{R}};{\bf{S}})$, 
where ${\bf{R}}=\left\{r_1\dots r_N\right\}$ (${\bf{S}}=\left\{s_1\dots s_N\right\}$) indicates the coordinates of the $N$ particles (shadow auxiliary variables) and $G({\bf{R}};{\bf{S}}) = \prod_{i<j}\phi_P(|r_i - r_j|)\prod_{i<j}\phi_S(|s_i - s_j|) \prod_{i} f_{PS}(|r_i - s_i|)$.
$f_{PS}(r)=\exp{(-C r^2)}$ is a Gaussian coupling between particle and shadow variables, while $\phi_{P,S}(r)=\exp{(-u_{P,S}(r)/2)}$ are Jastrow factors. We take $u_P(r)=(b_P/r)^{m_P}$ of the standard McMillan type, while $u_S(r)=(b_S/r)^{m_S} - \alpha \log{(\sin{(\pi r/L)})} \tanh{(r/\bar{r})}$, where $L$ is the size of the simulation box. The additional term in the Shadow factor is a long-range correlation of the one-dimensional
Reatto-Chester form [37]
, introduced to ensure a more rapid convergence of long-range correlations which are relevant for Luttinger liquids. In fact $\alpha$ can be related to the Luttinger 
parameter by $\alpha=2/K_L$. The $\tanh{(r/\bar{r})}$ factor is introduced solely to avoid perturbation of the short-range regime. All parameters in the trial wave-function are variationally optimized before projecting with the PIGS method, in order to maximize efficiency of the simulations. 
\begin{figure}[btp]
\begin{center}
\includegraphics[width=8.5cm]{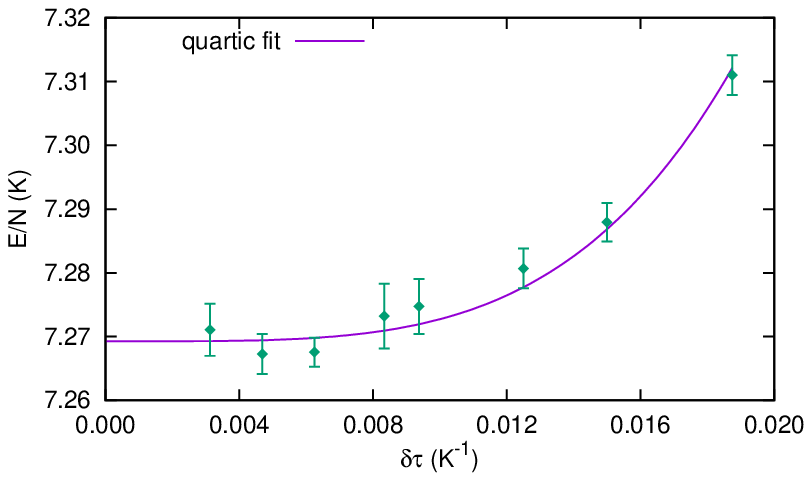}\\
\includegraphics[width=8.5cm]{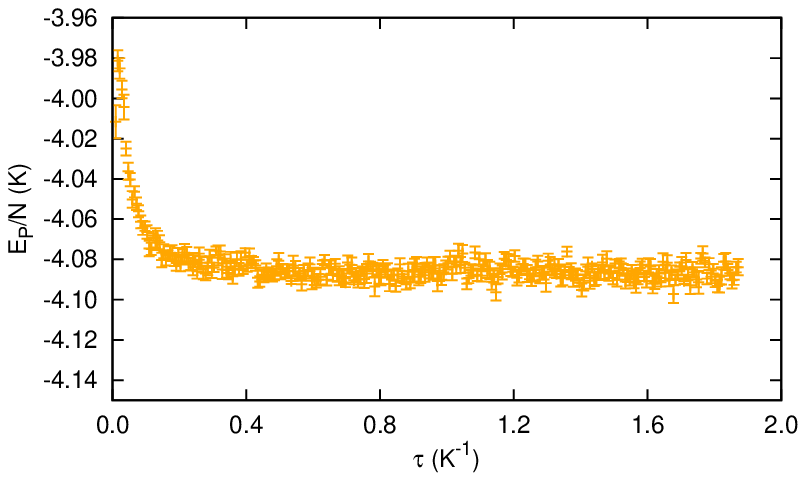}
\caption{Upper panel: Time-step analysis of the energy per particle of $N=10$ atoms at the density $\rho = 0.300 \text{\AA}^{-1}$, with a total small projection time of $\tau=0.075 K$. Solid line: quartic fit to the data. Lower panel: mixed potential energy of $N=100$ atoms at the density $\rho = 0.300 \text{\AA}^{-1}$ as a function of projection time. The corresponding variational estimate is $E_P/N=-4.377(6) K$.
}
\label{fig:dt}
\end{center}
\end{figure}

In the PIGS method the imaginary time $\tau$ of propagation is split into $M_P$ time steps of size $\delta\tau=\tau/M_P$ so that a suitable short-time approximation for the propagator can be used. We employ the fourth-order pair-Suzuki approximation [36]
 and observe convergence of ground-state estimates with a typical projection time $\tau \simeq 0.8\text{K}^{-1}$ using a time 
step $\delta\tau = 1/160\text{K}^{-1}$ (See Fig.~\ref{fig:dt} for typical time-step and total-time analyses). Once convergence is obtained, a further projection time of typical duration $\tau_F=M\delta\tau\simeq 2\text{K}^{-1}$ is used to sample the intermediate scattering function $F(q,\tau)$ (Eq. 4 in main text).
This projection time allows us to resolve the main features of the density fluctuations spectrum and, in the high-density regime, to quantitatively extract information on the spectral shape at low-energy. 

\section{Finite-size effects}

The vast majority of QMC methods give access to the properties of 
finite systems, made of $N$ particles enclosed in a spatial region
of volume $V$.
However, we are interested in the properties of the system in the limit $N \to \infty$ keeping the
density $\rho \equiv \frac{N}{V}$ fixed. We therefore must assess for which 
size results are compatible with those at the thermodynamic limit
within the statistical uncertainties of the simulation.

In this Section, we present and describe the size effects on the
equation of state, static structure factor and imaginary-time
correlation functions of up to $160$ Helium atoms at the highest
density $\rho = 0.300 \text{\AA}^{-1}$ to show that, except in special
circumstances, results for a system of $N=50$ Helium atoms are
representative of the thermodynamic limit. Nonetheless, static properties presented in this work have been calculated using up to $N=160$ particles.

\subsection{Equation of state}

In Table \ref{tab:eos0300} we report the ground-state energy per particle
$\mathcal{E}_N$ as a function of the number of atoms, at the density $\rho = 0.300 \text{\AA}^{-1}$.
The dependence of the results on $N$ is well captured by the
formula $\mathcal{E}_N = e_1 + \frac{e_2}{N^2}$ with $e_1 = 7.408(5) \, K$. 
This functional form reveals that size-effects on
the equation of state are modest, in that 
the ground-state energies of $40$ or more atoms are compatible 
with each other and with $e_1$. The expression of $\mathcal{E}_N$ also reflects the similarity 
between Helium atoms in the high-density regime and hard rods:
indeed, for the HR system, the relation $\mathcal{E}_N = e_1 
\left( 1- \frac{1}{N^2} \right)$ holds exactly [5].
For a more detailed comparison, we also report the ground-state energy per particle of systems at $\rho = 0.220 \text{\AA}^{-1}$. For $N \geq 30$, results are compatible with each other within the
statistical uncertainties of the simulations, around $2 \%$.
\begin{table}
\begin{tabular}{ c c c | c c c}
  \hline
  & $\rho = 0.300 \text{\AA}^{-1}$ & & & $\rho = 0.220 \text{\AA}^{-1}$ & \\
N & $\mathcal{E}_N \, [K]$ & $\sigma_N \, [K]$ & N & $\mathcal{E}_N \, [K]$ & $\sigma_N \, [K]$\\
  \hline
10  &  7.26   &    0.01   &   10  &  0.658   &    0.003 \\
20  &  7.37   &    0.01  &   20  &  0.673   &    0.002 \\
30  &  7.37   &    0.01  &   30  &  0.679   &    0.002 \\
40  &  7.41   &    0.01  &   40  &  0.679   &    0.002 \\
50  &  7.40   &    0.008  &   80  &  0.679   &    0.002 \\
80  &  7.41   &    0.008  &   160 &  0.682   &    0.002 \\
100 &  7.41   &    0.008  &       &          &          \\
160 &  7.41   &    0.009  &       &          &          \\
  \hline  
\end{tabular}
\caption{Equation of state at high density.} \label{tab:eos0300}
\end{table}

At lower densities, size effects on the equation of state are 
even more modest, implying that the ground-state energies of $20$ 
or more atoms are compatible with each other. It is worth recalling that $K_L$ can be estimated from the equation of state by fitting 
the latter to a polynomial in the density and computing $K_L$ as function of the fitting 
parameters.
Such procedure, however, involves error propagation from the fitting parameters to the 
estimator of $K_L$. At low density, this is responsible for the large error bars in the
estimates of $K_L$ in Fig.~1 of the main text.

\subsection{Static structure factor}

\begin{figure}[btp]
\begin{center}
\includegraphics[width=8.5cm]{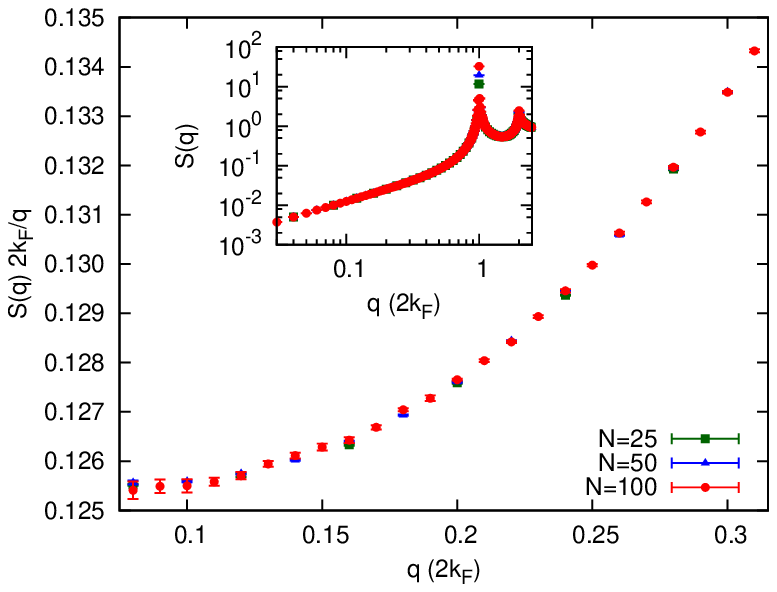}
\caption{Static structure factor at $\rho = 0.220 \text{\AA}^{-1}$ for various systems sizes (inset), divided by $q/2k_F$ at low momenta (main figure).
}
\label{fig:sq0300}
\end{center}
\end{figure}

In Fig.~\ref{fig:sq0300} we show the ratio $2 k_F S(q)/q$, for systems of $N=25,50$ and $100$ atoms at the highest density.
This quantity has been used to estimate the Luttinger parameter, taking advantage of the
relation $S(q) \sim K_L \, \frac{q}{2k_F}$, holding in the low-momentum regime.
Using results for $N=50$ particles, we obtain $K_L = 0.1255(5)$. This result is compatible
with that obtained for $N=100$ particles, confirming the robustness and the absence of
size effects of our estimate for $K_L$.

Away from a narrow neighborhood of the Umklapp points $q_j = 2 \, j \, k_F$, with $j$ 
integer, results for $S(q)$ relative to $N=50,100$ particles are compatible with each 
other within the statistical uncertainties of the calculations, which in all cases are well below 1 $\%$.
The presence of peaks diverging with the system size limits the possibility 
of producing estimates of $S(q_n)$, $S(q_n,\omega)$ free from size effects.
However, away from the Umklapp points, estimates of $S(q)$ for $N=50$ particles are
representative of the thermodynamic limit.

\subsection{Imaginary-time correlation functions}

\begin{figure}[ptb]
\begin{center}
\includegraphics[width=8.5cm]{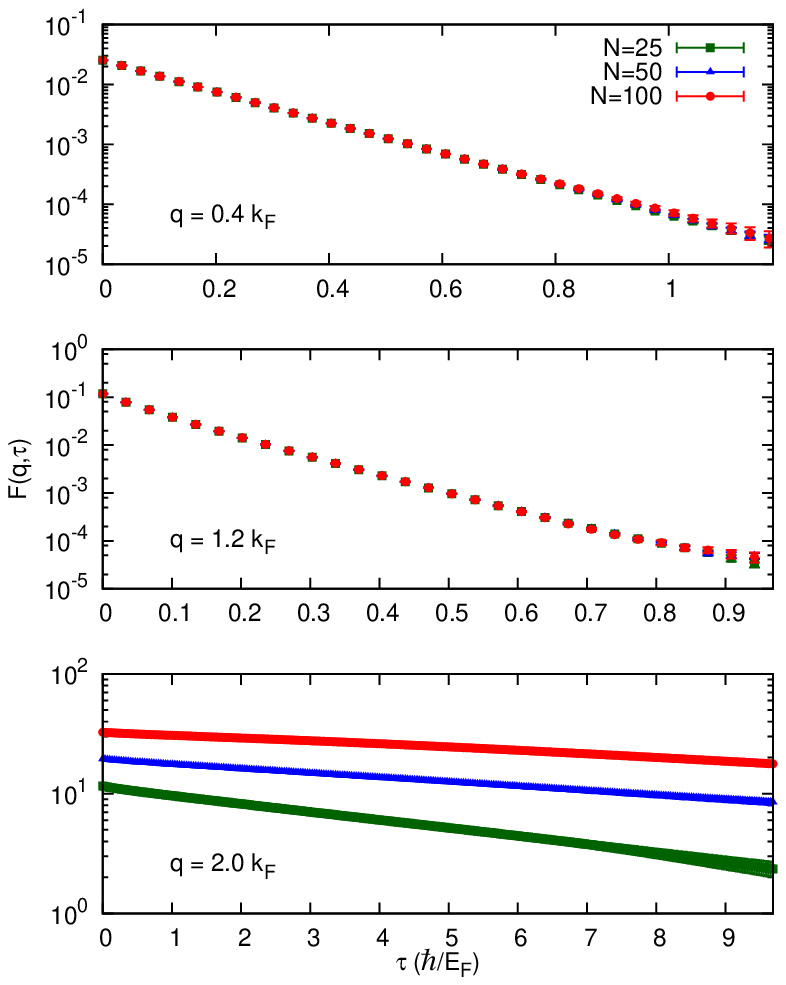}
\caption{(color online) 
Imaginary-time density-density correlation function $F(q,\tau)$ at $\rho = 0.300 \text{\AA}^{-1}$ for various systems sizes at three representative momenta.
}
\label{fig:ftau0300}
\end{center}
\end{figure}

Given the limited size effects observed in the static structure factor, and the central role 
of $S(q,\omega)$ in the present work, it is very important to investigate the size effects on 
the imaginary-time correlation functions. We illustrate $F(q,\tau)$ in Fig.~\ref{fig:ftau0300} for systems
of $N=25,50,100$ particles at the highest density $\rho = 0.300 \text{\AA}^{-1}$, for the wavevectors 
$q/k_F= 0.4, 1.2, 2$, representative of the low-momentum, intermediate-momentum and Umklapp regimes.
Size effects on imaginary-time correlation functions agree with those in the static structure
factor. In particular strong effects are seen only around the Umklapp points.

The analysis of the high-imaginary time region reveals that a precise determination of the 
low-energy threshold $\omega_\soglia(q)$ is made difficult by the growth of relative errors with imaginary time,
more than by size effects.
The uncertainty on the low-energy threshold naturally has a negative impact on the possibility
of assessing a power-law behavior of $S(q,\omega)$ close to $\omega_\soglia(q)$.
Therefore, in order to produce a quantitative estimate of the exponents of the nonlinear-TLL 
theory as presented in the main text, we employed analytical information for an equivalent hard rods system as described in the following
Sections.

\section{Genetic Inversion via Falsification of Theories method}
Eq. (4) in the main text is a Fredholm equation of the first kind and is an ill-conditioned problem, because a small variation in the imaginary-time intermediate scattering function $F$ produces a large variation in the dynamical structure factor $S$. At fixed momentum $q$, the computed values $F_j=F(q,j\delta\tau)$, where $j=0\dots M$, are inherently affected by statistical uncertainties $\delta F_j$, which hinder the possibility of deterministically infer a single $S(q,\omega)$, without any other assumption on the solution. The Genetic Inversion via Falsification of Theories method (GIFT) exploits the information contained in the uncertainties to randomly generate $Q$ compatible instances of the scattering function $F^{(z)}$, with $z=1\dots Q$, which are independently analyzed to infer $Q$ corresponding spectra $S^{(z)}$, whose \emph{average} is taken to be the ``solution''. This averaging procedure, which typifies the class of stochastic search methods [40-42],
yields more accurate estimates of the spectral function than standard Maximum Entropy techniques. Although any numerical analytic continuation method, included the GIFT method, is not able to precisely resolve multiple narrow peaks if they are present, apart from the lowest energy one, the most relevant features of the spectrum are retrieved in their position and (integrated) weight. In this work we show that with GIFT even some properties of the shape of the spectra close to the lowest threshold can be reliably inferred, once quite heavy reconstructions are performed.

Given an instance $z$, the procedure of analytic continuation from $F^{(z)}$ to $S^{(z)}$ relies on a stochastic genetic evolution of a population of spectral functions of the generic type $S^{(z)}(q,\omega) = c_0\sum_{i=1}^{N_\omega}s_i\delta(\omega-\omega_i)$, where $c_0=F^{(z)}(q,0)$ and the zeroth momentum sum rule $\sum_{i=1}^{N_\omega}s_i=1$ holds. The $N_\omega$ support frequencies $\omega_i=\omega_\soglia + \Delta\omega(i-1/2)$ are spaced by a small $\Delta\omega$ and a minimum threshold frequency $\omega_\soglia$ can be imposed. Genetic algorithms provide an extremely efficient tool to explore a sample space by a non-local stochastic dynamics, via a survival-to-compatibility evolutionary process
mimicking the natural selection rules; such evolution aims toward increasing the \emph{fitness} of the individuals, defined as
\begin{multline}
\Phi^{(z)}(S) =
-\sum_{j=0}^M \frac{1}{\delta F_j^2}\left[F^{(z)}_j - c_0\sum_{i=1}^{N_\omega} e^{-j \delta\tau\omega_i}\,s_i\right]^2
\\
- \gamma \left[c_1 - c_0\sum_{i=1}^{N_\omega} \omega_i\,s_i\right]^2 \;,
\label{fitness}
\end{multline}
where the first contribution favors adherence to the data, while the second one favors the fulfillment of the f-sum rule, with $c_1=\frac{\hbar q^2}{2m}$ and $\gamma$ a parameter to be tuned for efficiency. A step in the genetic evolution replaces the population of spectral functions with a new generation, by means of the ``biological-like'' processes of {\it selection}, {\it crossover} and {\it mutation}, which are described in detail in [43].
In this work we have added a \emph{smoothening mutation}, which operates randomly on very small portions of the spectrum, and a \emph{long-range mutation} which exchanges weight between two randomly separated bins. Moreover, the genetic evolution is tempered by an acceptance/rejection step based on a reference distribution $p^{(z)}(S)=\exp{(\Phi^{(z)}(S)/T)}$, where the coefficient $T$ is used as an effective temperature in a standard simulated annealing procedure [40]. 
We found that this combination is optimal in that it combines the speed of the genetic algorithm with the prevention of strong mutation-biases thanks to the simulated annealing.
Convergence is reached once $|\Phi^{(z)}(S)|< 1$, and the best individual, in the sense that it does not falsify the theory represented by \eqref{fitness}, is chosen as the representative $S^{(z)}$. The final spectrum, as in Fig.~3 of the main text (where no threshold is assumed), is obtained by taking the average over instances $\bar{S}(q,\omega)=\frac{1}{Q}\sum_{z=1}^Q S^{(z)}(q,\omega)$.

Note that the space of spectral functions that we consider is quite general, and can be extended by using smaller frequency spacing $\Delta\omega$. However, the purpose of GIFT is not to exactly resolve the spectrum of a finite system, which would be indeed consisting of a sum of delta-functions, but to provide a convolution, which, as such, is suited to study the thermodynamic limit. The proper quantity to be analyzed to define such limit is in fact $F(q,\tau)$, whose size effects we have described in the previous Section.

\begin{figure}[ptbh]
\begin{center}
\includegraphics[width=8.5cm]{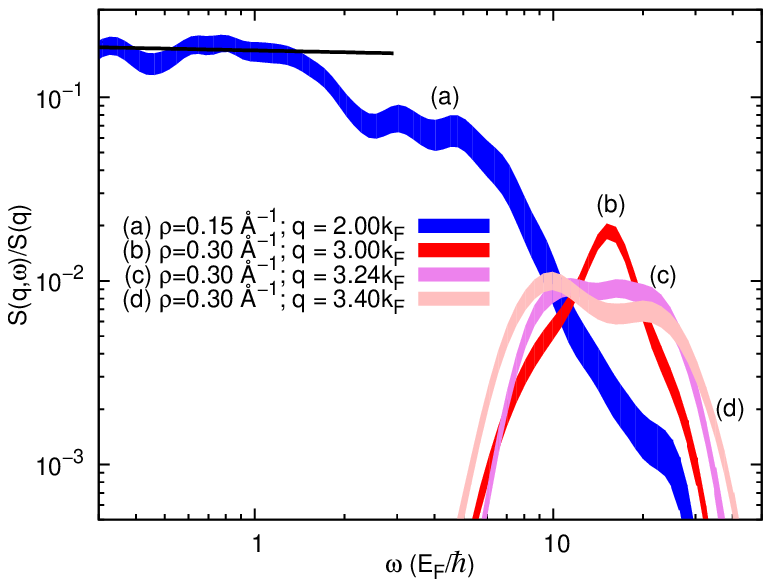}
\caption{$S(q,\omega)$ as a function of frequency for different momenta and densities, divided by the corresponding $S(q)$. Data are post-processed with a smoothening operation, while the width indicates the typical variance of reconstructions. Solid line: see text.}
\label{fig:exampleq}
\end{center}
\end{figure}

\emph{Dynamical Structure Factor at specific momenta.} In Fig.~\ref{fig:exampleq} we show some reconstructed spectra at specific momenta, with the aim of highlighting the appearance of regions of almost flat spectral weight, either at $q=2k_F$ for small $\omega$ when $K_L\simeq 1$, or around the special ${\cal Q}_1 = 2\pi/a$ at high density. In particular curve (a) corresponds to density $\rho = 0.150 \text{\AA}^{-1}$, for which $K_L\simeq 0.98$. Data are shown above the minimal frequency corresponding to the super-current state $\hbar\omega=(2 \hbar k_F)^2/2m N$. The comparison to the TLL prediction $S(2k_F,\omega) \propto \omega^{2(K_L - 1)}$ [62-64]
(solid line) is quite satisfactory. Curves (b-d) show the spectra at the density $\rho = 0.300 \text{\AA}^{-1}$ around $q=3.24 k_F=3.05\text{\AA}^{-1}$, which is the momentum displaying maximally flat spectrum in our reconstruction. The spectra shown in Fig.~\ref{fig:exampleq} are representative of the data used to produce Fig.~3 in the main text, where no threshold is assumed. More sophisticate and accurate continuations are used to produce Fig.~4 in the main text, with the procedure described in the next section.

\section{Power-law fit of the reconstructed spectra}
We now describe with some detail the procedure employed to obtain the results presented in Fig.~4 of the main text.

In order to remove noise, we assume that the spectral weight is zero below the threshold of a system of $N=50$ Hard-Rods whose radius is fixed so as to have the same $K_L=0.1255$ as for our $^4$He system at the density $\rho=0.3\text{\AA}^{-1}$. This is the only additional knowledge about $S(q,\omega)$ enforced in our reconstruction.
We take advantage of the following analytical expression for the threshold of a system of $N$ hard rods:
\begin{equation}
 \omega_\soglia^{(N)}(q) = \omega_\soglia(q)\left(1+\frac{1}{N}\right) + \omega_{sc}(q) \;,
\end{equation}
where the threshold in the thermodynamic limit is $\omega_{\soglia}(q)=\omega^-(q-2n k_F)/K_L$ with $2 n k_F < q < 2(n+1) k_F$ and $n$ integer, as described in the main text, while two corrections have to be added for finite systems: $\omega_{\soglia}(q)/N$ and $\omega_{sc}(q)=\frac{\hbar q^2}{2m N}$, corresponding to the energy of a super-current state [76]. 
When the threshold cannot be reasonably assumed, a general method has been recently proposed [79].
In the future, it would be interesting to investigate whether such method can yield reliable threshold energies in the intermediate-to-low-energy regime. 

\begin{figure}[!h]
\begin{center}
\includegraphics[width=8.0cm]{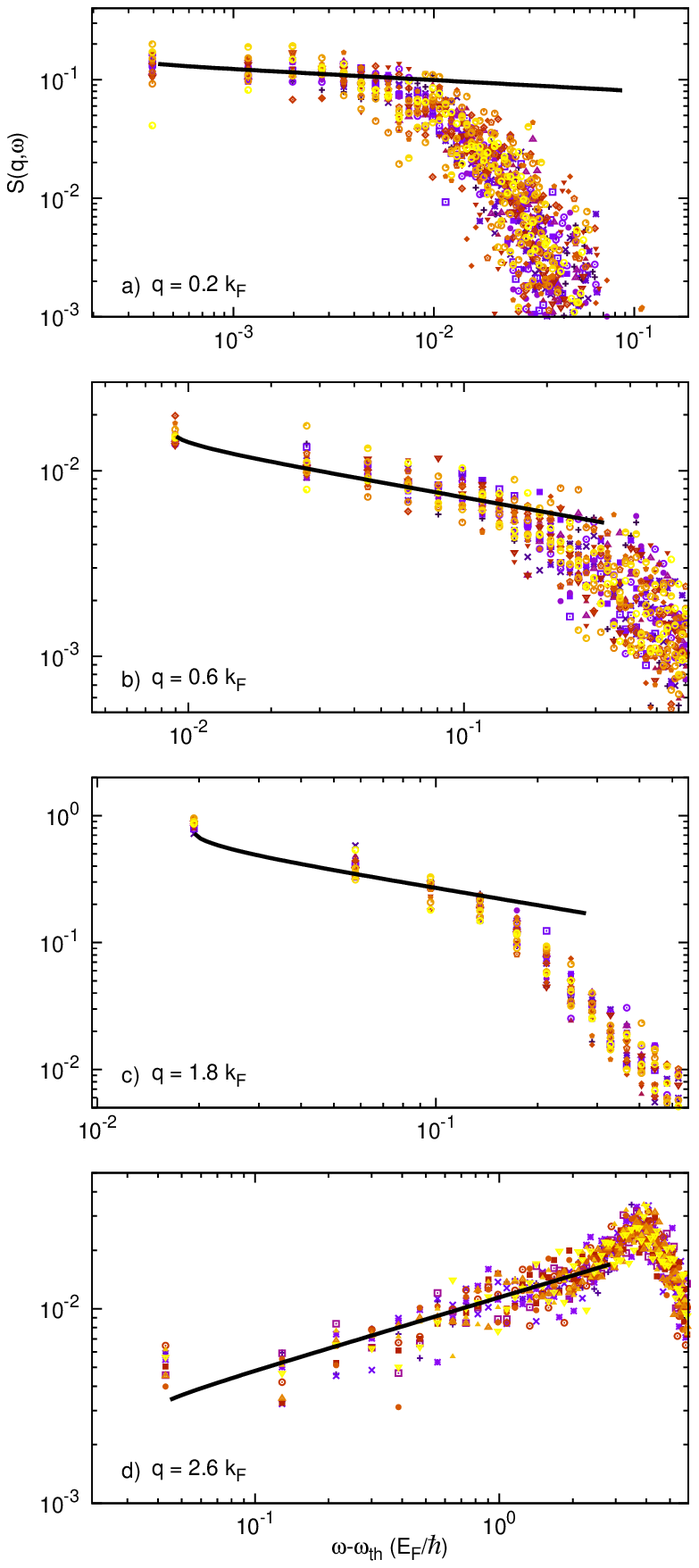}
\caption{Set of block averages of the dynamical structure factor reconstructed at $\rho = 0.300 \text{\AA}^{-1}$ and various momenta. Solid lines are Eq.\eqref{eq:blockmodel}, using the mean of $A$ and $\mu$ obtained from the fits of the block averaged spectra. Panel a: momentum $q=0.2k_F$ and $\hbar\omega_\soglia=2.927E_F$. Panel b: $q=0.6k_F$ and $\hbar\omega_\soglia=6.834E_F$. Panel c: $q=1.8k_F$ and $\hbar\omega_\soglia=2.991E_F$. Panel d: $q=2.6k_F$ and $\hbar\omega_\soglia=6.962E_F$.}
\label{fig:sqw-fit}
\end{center}
\end{figure}

For each momentum, we then exploit the imaginary-time correlation functions $F(q,\tau)$ obtained with the PIGS simulations, with their statistical error bars, to sample $Q=1280$ compatible $F^{(z)}$. Instead of taking a single average, as in the previous section, the $Q$ reconstructed spectra are averaged over blocks of 128 elements, yielding $B=10$ estimated spectra, which are fitted with the following expression:
\begin{equation}
 S_i=A (\omega_i-\omega_\soglia)^{-\mu}  Q_i\;,\label{eq:blockmodel}
\end{equation}
where $Q_i=\frac{\tilde{\omega}_i}{(1-\mu)}\left[\left(1+\frac{1}{2\tilde{\omega}_i}\right)^{1-\mu}-\left(1-\frac{1}{2\tilde{\omega}_i}\right)^{1-\mu}\right]$ with $\tilde{\omega}_i~=~(\omega_i-\omega_{th})/\Delta\omega$. This expression is the average value of the power-law model in the interval $[\omega_i-\Delta\omega/2,\omega_i+\Delta\omega/2]$, and it is the correct way to compare our discrete-frequency spectra to the model, especially close to $\omega_\soglia$ when a divergence is expected. The range of the fitting procedure is 
from $\omega=\omega_\soglia$ to the maximum frequency yielding a reduced Chi-square $\chi^2/\nu\simeq 1$. Such maximum frequency for a best fit depends on $q$ but it is typically of order $q\approx 0.3 k_F$. Each of the $B$ fits yields a value for $\mu$, the mean value and standard deviation of which are showed in Fig.~4 of the main text. 
In Fig \ref{fig:sqw-fit} we show some of the averaged spectra at different momenta together with Eq. \eqref{eq:blockmodel} using the mean of $A$ and $\mu$. In panel (d) one can see that the first frequency bin is not well fitted by the power-law model. This is typical of the spectra where the spectral weight decreases towards the threshold. The effect is much more pronounced for the momenta close but higher than multiples of $2k_F$, which has prevented us to extract reliable exponents in such cases. We cannot assess whether this indicates a truly finite spectral weight, beyond the power-law model, or, more probably, an unavoidable limitation of analytical continuation strategies. However, for negative values of the exponent, giving rise to a divergence in the spectrum at $\omega=\omega_\soglia(q)$, our procedure has provided robust evidence of a power-law singularity in the dynamical structure factor consistent with the HR model.
\vfill
\pagebreak

\vspace{1cm}
\vspace{1cm}

\end{document}